\documentclass[review,twocolumn]{elsarticle}

\usepackage{lineno,hyperref}
\modulolinenumbers[5]

\journal{Journal of Electron Spectroscopy and Related Phenomena}

\begin{document}

\begin{frontmatter}
\title{Comment on: Quantitative study of the $f$ occupation in Ce$M$In$_5$ and other cerium compounds with hard X-rays}
\author[myaddress]{L.~Howald}
\ead{ludovic.howald@psi.ch}
\address[myaddress]{Swiss Light Source, Paul Scherrer Institut, CH-5232 Villigen PSI, Switzerland}
\end{frontmatter}

The valences of cerium dioxide (CeO$_2$) and cerium fluoride (CeF$_4$) have long been a subject of controversy. However the controversy is more related to different nomenclature conventions than to different physical interpretations. Indeed, presenting their photoemission and deep core-level spectroscopies results, Kaindl \textit{et al.} conclude: ``This approach confirms that CeF$_4$ is tetravalent, but does not deny the covalent admixture of 4$f$ character in the valence band.'' \cite{Kaindl1987}, followed by a later comment ``The purpose of our Letter was to provide a simple experimental criterion to distinguish between mixed-valent (metallic) Ce compounds and the wide-gap, tetravalent insulators CeF$_4$ and CeO$_2$, which have been and are still classified by several workers as mixed valent.'' \cite{Kaindl1988}.  In contrary using an Anderson model Kotany \textit{et al.} obtained: ``The conclusion of the present paper is as follows:(a) The ground state of CeO$_2$ is the mixed valence state where 4$f^0$ and 4$f^1$ configurations are strongly mixed.[...] (c) The 4$f$ level $\epsilon^0_f$ is near the top of the valence band.'' \cite{Kotani1985} and propose a 4$f$ occupancy of 0.29 for CeF$_4$ and 0.5 for CeO$_2$  \cite{Kotani1987}, using notably data from Wuilloud \textit{et al.}, who had initially conclude: ``A mixed valence can be definitely excluded in CeO$_2$''\cite{Wuilloud1984}. The texts of these authors and other cited within these works, indicate that the term: ``valence'' is not uniquely defined. For Kaindl \textit{et al.} electrons of 4$f$ character in covalent bond are part of the valence. In contrary for Kotany \textit{et al.} electrons with 4$f^0$ and 4$f^1$ configurations are not part of the valence even if the strong hybridization and energy level $\epsilon^0_f$ within the valence band indicate an at least partial covalent bond formation.

For cerium we have to differentiate between the occupancy of the purely atomic 4$f$ orbital ($n_{fO}$) and the amount of electrons with 4$f$ character ($n_{fH}$) in hybridized orbitals. Both contributions behave as localized 4$f$ states. Hybridization can come from chemical bonds such as covalent bonds or formation of an electronic band of partial 4$f$ character, in contrary to valence bonds, valence band or ionic bonds, in which the 4$f$ character is lost or extended.
With their model Kotany \textit{et al.} can only obtained the total amount of electrons with 4$f$ character ($n_{fO}+n_{fH}$). The model requires a strong hybridization to reproduce the experimental results indicating that more than the purely atomic 4$f$ orbitals are considered. They define the valence of cerium as: $4-(n_{fO}+n_{fH})$. In contrary, experimentally, using susceptibility or a direct interpretation of x-ray photoemission or absorption spectra one directly obtains an approximate value of $n_{fO}$ and the valence can be defined as all electrons participating in the bonding: $4-n_{fO}$. For CeO$_2$ and CeF$_4$ the above mentioned literature results indicate that $n_{fO}\simeq 0$ while $n_{fH}\simeq 0.5$ or respectively $n_{fH}\simeq 0.29$. The valence value depends on its definition. Note that $n_{fO}$ can take non-integral values with dynamic fluctuations of electrons between the 4$f$ shell and other orbitals of the same atom \cite{Roehler1987}.

The same issue arises when looking at heavy fermions materials such as CeCoIn$_5$. Using an Anderson impurity model, Sundermann \textit{et al.} obtained a total 4$f$ electron count $n_{fO}+n_{fH}=0.97$ \cite{Sundermann2016}. In contrary the valence of 3.15 obtained by comparison of x-ray absorption spectra at the M-edge \cite{Howald2015} confirms the orbital occupancy obtained at the L-edge $n_{fO}\simeq 0.85$ \cite{Booth2011}. What is remarkable is that both obtained 4$f$ state occupancies ($n_{fO}+n_{fH}$ and $n_{fO}$) are unmodified between a magnetic system [CeRhIn$_5$ in Ref~\cite{Sundermann2016} and CeCo(In$_{0.85}$Cd$_{0.15}$)$_5$ in Ref.~\cite{Howald2015}] and the paramagnetic system CeCoIn$_5$. In heavy fermion materials, the occupancy $n_{fO}$ is the essential ingredient of the mass renormalization via the Kondo lattice, while for CeCoIn$_5$ it was recently demonstrated that the magnetism occurs within a band of partial 4$f$ character \cite{Howald2015}, namely of 4$f$ occupancy $n_{fH}$. The absence of variation in both $n_{fH}$ and $n_{fO}$ at the antiferromagnetic--paramagnetic transition is an extremely strong result as it demonstrates that the magnetic transition is independent of the Kondo physics. Indeed, the occupancy of the two independent subsets of 4$f$ electrons $n_{fH}$ and $n_{fO}$ responsible for magnetism and Kondo correlations is unaffected by the magnetic transition. 

\section*{References}

%\bibliography{C:/Users/howald_l/Documents/Publi/biblio}

\end{document}